\documentstyle[prl,aps,multicol]{revtex}
\begin{document}
\author{K. Kiran Kumar\thanks{%
permanent address: Department of Chemistry Indian Institute of Technology
Kharagpur India } and K.L. Sebastian}
\address{Department of Inorganic and Physical Chemistry\\
Indian Insitute of Science\\
Bangalore 560012\\
India}
\title{Adsorption assisted translocation of a chain molecule through a pore in a
spherical vesicle }
\date{August 28, 1999}
\maketitle

\begin{abstract}
We analyze the free energy for translocation of a polymer from the outside
of a spherical vesicle to the inside. The process is assumed to be driven by
the adsorption of the polymer on the inner surface of the vesicle. We argue
that in the case where the polymer is adsorbed on the outer surface too, the
entropic barrier for translocation is absent. We analyze the adsorption
energy and find the free energy profile for the process. We argue that the
motion corresponds to a polymer crossing a region with a change in free
energy per segment. Based upon our earlier analsis of the behaviour of kinks
in such a problem, we conclude that the translocation can occur with a
crossing time $t_{cross}\sim N$.
\end{abstract}

\section{\bf Introduction}

Translocation of a long chain molecule through a pore (see figure 1A) has
been a problem of current interest \cite{park1}, \cite{park2}. Park and Sung 
\cite{park1} have studied the problem of release of a polymer through a
small pore in a spherical vesicle. Their primary results concern the entropy
effects in driving the polymer from the inside to the outside of the
spherical vesicle. They calculate the time taken by a polymer molecule to
translocate through the pore. They have considered both the inner and the
outer surfaces of the vesicle to be hard walls, which repell the units of
the long chain molecule. They argue (a) that the membrane curvature drives
the polymer out of a spherical vesicle (b) capture of a polymer of $N$
segments into a a small bud takes a very long time proportional to $\exp (N)$%
, which can be reduced to $N^3$ or even to $N^2$ by free energy differences
between the inside and the outside. In this paper, we consider the case
where there may be adsorption on both the surfaces. We construct a free
energy profile for the translocation process. We show that in the case, the
entropic barrier for penetration in to the pore is absent. Following the
reference \cite{Sebastian}, we suggest that translocation can occur by kink
motion and this is driven by the free energy difference due to the different
adsorption strengths. This can lead to a translocation time $t_{cross}\sim N$
.

We make use of the approach of de Gennes \cite{de Gennes} to treat the
adsorption of the chain molecule on the surfaces \cite{eisen}. We go beyond
the results of Park and Sung \cite{park1} and find the exact partition
function for a chain restricted to the outside of a sphere and having an
attractive interaction with the surface of the sphere. For a chain molecule
confined to the inner region of the sphere and interacting with the surface,
we give the exact result for the Laplace transform of the partition
function. We analyze the condition for the existence of an adsorbed state
and find that they are different for the two cases. Partition function is
found in the limit of a long chain and this is used to calculate the free
energy of the adsorbed polymer. From this, we construct the free energy
profile for the translocation process.

\section{Green's functions for adsorption problems}

We wish to evaluate the partition function for a chain of $n$ segments
confined to the outer region or inner region of a sphere. For this, we
consider the Green's function $G({\bf r,r}_0;n)$, obeying the equation $(%
\frac \partial {\partial n}-\frac{b^2}6\nabla ^2+V(r))G({\bf r,r}%
_0;n)=\delta (n)\delta ({\bf r-r}_0)$ (see\cite{park1} for more details). $%
V(r)$ is the potential of interaction arising from the sphere. The partition
function for the polymer may be calculated from the angle averaged function $%
Z(r,r_0;n)=\int_0^\pi d\theta \int_0^{2\pi }d\phi \sin (\theta )G({\bf r,r}%
_0;n)$. This averaged Green's function obeys the simpler equation \cite
{park1}

\begin{equation}
\label{one}\left( \frac \partial {\partial n}-\frac{a^2}{r^2}\frac \partial {%
\partial r}r^2\frac \partial {\partial r}+V(r)\right) Z(r,r_0;n)=\frac 1{r^2}%
\delta (n)\delta (r-r^{^{\prime }}) 
\end{equation}
where $a=b/\sqrt{6}$. The conditions on $Z(r,r_0;n)$ are that $%
Z(r,r_0;n)\rightarrow 0\,$ as $r,r_0\rightarrow \infty $. In the case $%
V(r)=0 $, we can solve this equation easily to obtain

\begin{equation}
\label{two}Z_0(r,r^{^{\prime }};n)=\frac 1{rr_0a\sqrt{\pi n}}\exp \{-\frac{%
(r^2+r_0^2)}{4a^2n}\}\sinh (\frac{rr_0}{2a^2n}) 
\end{equation}

The presence of a spherical surface at $r=R$ with a short range attractive
(or replusive) potential can be accounted for by imposing the de Gennes'
boundary condition \cite{de Gennes} $\left( \frac{d\ln Z(r,r_0;n)}{dr}%
\right) _{r=R}=c_0$. If $c_0\rightarrow \infty $, the surface is a hard
wall. $c_0=0$ represents a neutral surface, while $c_0<0$ models an
attractive surface. We now use this to solve the problem of adsorption on
the outside of a sphere. It is easier to work with the Laplace transform of
the Green's function $\overline{Z}(r,r^{^{\prime }};s)=$ $\int_0^\infty
dnZ(r,r^{^{\prime }};n)e^{-sn}$. It obeys the differential equation

\begin{equation}
\label{three}\left( s-\frac{a^2}{r^2}\frac \partial {\partial r}r^2\frac 
\partial {\partial r}\right) \overline{Z}(r,r_0;s)=\frac 1{r^2}\delta
(r-r_0) 
\end{equation}
and the boundary condition becomes

\begin{equation}
\label{threea}\left( \frac{d\ln \overline{Z}(r,r_0;s)}{dr}\right) _{r=R}=c_0 
\end{equation}
.

\subsection{Adsorption on the outer surface of a sphere}

We now solve the equation (\ref{three}) in the region outside the sphere ($%
r>R$) subect to the de Gennes ' boundary condition to obtain $Z_{>}(r,r_0;s)$%
. We get

$Z_{>}(r,r^{^{\prime }};n)=$%
$$
\frac 1{2\,a\,\,\sqrt{\pi n}\,\,r\,r_0}\left( \exp [-\frac{(r-r_0)^2}{4a^2n}%
]+\exp [-\frac{(r+r_0-2R)^2}{4a^2n}]\right) - 
$$

\begin{equation}
\label{four}\frac c{r\,r_0}\,e^{a^2\,c^2\,n+c\,\left( r-2\,R+r_0\right)
}\,Erfc(a\,c\sqrt{n}+\frac{r-2\,R+r_0}{2\,a\,\sqrt{n}}) 
\end{equation}
where $c=(c_0+\frac 1R)$, and $Erfc$ is the complementary error function.
This is similar to the expression for the case of adsorption on a planar
surface \cite{park2}. As the wall becomes repulsive ($c\rightarrow \infty $%
), the above reduces to the solution for absorbing barrier condition:

\begin{equation}
\label{five}Z_{>}(r,r_0;n)=\frac{\exp [-\frac{(r-r_0)^2}{4a^2n}]-\exp [-%
\frac{(r+r_0-2R)^2}{4a^2n}]}{2arr_0\sqrt{\pi n}} 
\end{equation}

Note that in the limit $R\rightarrow 0$ the parameter $c\rightarrow \infty $
and the result is the free space solution of equation (\ref{two}).

\subsection{Partition function for a polymer with one end on the surface}

We find the partition function for the polymer with the end $r_0$ fixed at
the surface of the sphere $(r_0=R)$ and the other end free. This is found by
integrating over all the values of $r$. That is, $Q_{>}(n)=\int_R^\infty
drZ_{>}(r,R;n)r^2$. The result is

\begin{equation}
\label{six}Q_{>}(n)=\frac 1{c\,R}\left( 1+e^{a^2\,c^2\,n}\,\left(
-1+c\,R\right) \,Erfc(a\,c\,\sqrt{n})\right) 
\end{equation}
We now consider the limit where $n\rightarrow \infty $. If $c<0$, then we
can approximate the above as $Q_{>}(n)\sim \frac 2{R\,c}\left(
\,R\,c-1\right) \,e^{a^2\,n\,c{}^2}$. In this limit, the free energy of a
polymer, having $n$ units, with one end at $R$ is $F(n)=-k_BT\ln
Q_{>}(n)\simeq -k_BTa^2\,n\,c{}^2$. This gives a free energy per unit equal
to $-k_BTa^2\,\,c{}^2$ for the chain on the outer surface of the sphere.
This free energy primarily comes from the bound (adsorbed) state on the
surface of the sphere. We can also ask: what is the condition for the
existence of the bound state? Obviously, it would be there, if $c<0$. In
contrast to the situation on a planar surface where the condition is $c_0<0$%
, in this case, the attractive interaction has to below a finite critical
value - that is $c_0<-1/R$, to have a bound state.

\subsection{Adsorption on the inner surface of the sphere}

For the case where the polymer is inside the vesicle the boundary condition
we impose is $\left( \frac{d\ln \overline{Z}(r,r_0;s)}{dr}\right)
_{r=R}=-d_0 $. Note that we have put a negative sign in front of $d_0$
(compare the equation (\ref{threea})) as increasing the value of $r$ means
one is moving towards the surface). One can explicitly evaluate the Laplace
transform of the propagator. On solving the equation (\ref{three}) we find

$\overline{Z}_{<}(r,r_0;s)=\frac{-\sinh (\frac{\sqrt{s}\,\left( r-r_0\right) 
}a)}{a\,r\,\,r_0\sqrt{s}}+$

$$
\frac{\sinh (\frac{r\,\sqrt{s}}a)}{a\,r\,\,r_0\sqrt{s}}\left( \frac{R\,\sqrt{%
s}\,\cosh (\frac{\left( R-r_0\right) \,\sqrt{s}}a)-a\,\sinh (\frac{\left(
R-r_0\right) \,\sqrt{s}}a)+a\,R\,\sinh (\frac{\left( R-r_0\right) \,\sqrt{s}}%
a)\,d_0}{R\,\sqrt{s}\,\cosh (\frac{R\,\sqrt{s}}a)-a\,\sinh (\frac{R\,\sqrt{s}%
}a)+a\,R\,\sinh (\frac{R\,\sqrt{s}}a)\,d_0}\right) 
$$

if $r>r_0$ and

$$
\overline{Z}_{<}(r,r_0;s)=\frac{\sinh (\frac{r\,\sqrt{s}}a)}{a\,r\,r_0\,%
\sqrt{s}}\,\left( \frac{R\,\sqrt{s}\cosh (\frac{\sqrt{s}\,\left(
R-r_0\right) }a)+a\,\sinh (\frac{\sqrt{s}\,\left( R-r_0\right) }a)\,\left(
-1+R\,d_0\right) }{R\,\sqrt{s}\,\cosh (\frac{R\,\sqrt{s}}a)+a\,\sinh (\frac{%
R\,\sqrt{s}}a)\,\left( -1+R\,d_0\right) }\right) 
$$

if $r<r_0.$

\subsection{Partition function for the polymer with one end on the inner
surface}

We find the partition function for the polymer with the one end fixed at the
inner surface of the sphere and the other end free. This is done by putting $%
r_0=R$ and integrating over all the values of $r$ inside the sphere. We get

$\overline{Q}_{<}(s)=\int_0^Rr^2dr\overline{Z}_{<}(r,R;s)$

$$
{=}\frac{\sqrt{s}-\tanh \left( R\sqrt{s}/a\right) a/R}{s\left( \sqrt{s}%
+ad\tanh \left( R\sqrt{s}/a\right) \right) } 
$$
where $d=d_0-1/R$. We can evaluate the partition function by Bromwich
integration: $Q_{<}(n)=\int_{\gamma -i\infty }^{\gamma +i\infty }dse^{sn}%
\overline{Q}(s)/(2\pi i)$ , where $\gamma $ is taken such that it is to the
right of all the poles of $\overline{Q_{<}}(s)$. To find the poles of $%
\overline{Q}(s)$, we solve the transendental equation $\tanh (x)=-x/(Rd)$,
where $x=R\sqrt{s}/a$. We consider the two different possibilities:

\begin{enumerate}
\item  $d_0$ is negative. Then $Rd<-1.$ Then one pole exists for real,
positive value of $s$. This value of $s$ is given by $s=\left( ax_r/R\right)
^2$, where $x_r$ obeys $\tanh (x_r)=-x_r/(dR)$. For this case, for large
values of $n$, the major contribution to $Q_{<}(n)$ comes from this pole and
hence $Q_{<}(n)\sim e^{\left( ax_r/R\right) ^2n}$and the free energy of the
adsorbed chain is $-k_BT\left( ax_r/R\right) ^2n$. The free energy per unit
length of the adsorbed chain is $-k_BT\left( ax_r/R\right) ^2$. In contrast
to the outer surface, there is no critical value for the logarithmic
derivative, for an adsorbed state to exist. It would exist as long as the
surface-segment interaction is attractive.

\item  $d_0\,$ is positive. In this case, all the poles have $s<0$ and there
is no adsorbed state. There are several states inside the spherical vessel,
and the Bromwich integration leads to an infinite sum for the partition
function. This type of problem has already been considered in \cite{park1}
and we shall not discuss this case further.
\end{enumerate}

\section{\bf Free Energy for the translocation process}

Let us now consider the translocation of the chain molecule. If the molecule
is confined to a spherical vesicle, with the outer and inner walls of the
vesicle having no affinity to the units of the chain molecule, then the free
energy of the molecule inside would be greater than on the outside.
Consequently, if there is a pore on the wall of the vesicle, the molecule
would move from the inside to the outside. But there are examples where the
molecule does the reverse \cite{Alberts}\cite{Peskin}. This would either
require either a motor driving the chain in, or a situation where the chain
has a lower free energy inside the vesicle. We consider the latter
situation, and we assume that the polymer can adsorb on inner walls of the
vesicle. (It is not necessary that the adsorption should be on the walls, it
can be anywhere inside the vesicle). The process that we consider is given
in the figure 1A. We assume that the strength of adsorption on the inner
wall is greater, and this drives the translocation process. The
translocating chain may be thought of as divided in to three portions -
first is the portion adsorbed on the outside, second, the portion on the
inside, and the third is the portion inside the pore. We already have
expressions for the free energy per segment of the chain molecule when it is
inside/outside. The portion that is inside the pore, is not adsorbed
anywhere and consequently, the free energy per segment is higher. Hence, the
free energy per segment for a translocating chain would have the appearence
of figure 2. An alternate scenario would be to have a pore which has
affinity towards the molecule, as a result of which the free energy per
segment follows the dotted line in the figure 2, for segments inside the
chain. Irrespective of which is the profile, if one now uses a one
dimensional Rouse model to describe the process \cite{Sebastian}, then one
has the following picture: the portion inside the pore is to be thought of
as a kink on the chain, and it can move on the chain. In the case where
there is a free energy lowering on going to the inner side, the kink will
move in the opposite direction with a finite velocity (for an expression for
this velocity, see \cite{Sebastian}). Consequently, the polymer will move in
to the inner side, with a constant rate. This means that the crossing time $%
t_{cross}\sim N$, in contrast to $t_{cross}\sim N^2$ or $t_{cross}\sim N^3$,
found in reference \cite{park1} It is of interest to consider the case where
the polymer is not adsorbed on the outer surface of the sphere, as in figure
1B. In this case, having one end of the polymer inside the pore would be
entropically unfavorable and would lead to a barrier, purely from this (see
for example, the figure 2 of reference \cite{park1} ). This type of barrier
is absent in our case. Thus, weak adsorption on the outer surface eliminates
the entropic barrier and thus facilitates translocation in to the vesicle.

One can also consider the case of translocation from a vesicle of radius $%
R_1 $to another of radius $R_2$. Analyzing this case, Park and Sung \cite
{park1} conclude that spontaneous capture in to a small bud can only rarely
occur as the chain is losing entropy in going in to a small bud. However,
adsorption on the surface of the bud, can drive the process. Park and Sung
suggest that this will lead to at the most $t_{cross}\sim N^2$. Motion of
the kink can act as the mechanism of transfer and this can lead to capture
times proportional to $N$.

\underline{{\bf Acknowledgements}}

K.Kiran Kumar is greatful to Prof.A.B. Sannigrahi for his encouragement in
his pursuit of science. He is thankful to Indian\ Academy of Science for
financial support in form of a Summer Associateship, which enabled him to
visit the Indian Institute of Science, Bangalore, where the work was done.

\subsection{Figure Captions}

\begin{enumerate}
\item  \begin{center}
Figure 1A: Case where there is adsorption on the inside as well as outside.
The translocation from the outside to the inside. The portion of the chain
within the pore is indicated by enclosing it within a circle. Figure 1B:
Case where there is no adsorption on the outside.

\item  Figure 2: The Free energy per segment as a function of segment
position.
\end{center}
\end{enumerate}


\begin{references}
\bibitem{park1}  P.J.Park,W.Sung, Phy Rev E57,730(1998)

\bibitem{park2}  P.J.Park,W.Sung, J.Chem.Phys,108,3013(1998)

\bibitem{Sebastian}  K.L. Sebastian, cond-mat/9907003, LANL reprint archive
(1999).

\bibitem{de Gennes}  P.G. de Gennes, Scaling Concepts in Polymer Physics,
Cornell University Press, Ithaca (1979).

\bibitem{eisen}  E.Eisenreigler,Polymers near surfaces,World Scientific
Publishers, Singapore (1993).

\bibitem{parkprl}  P.J.Park,W.Sung,Phys.Rev.Lett,77,783,(1996)

\bibitem{Alberts}  B. Alberts, D. Bray, A. Johnson, J. Lewis, M. Raff, K.
Roberts and P. Walker, Essential Cell Biology, Garland Publishing Inc, New
York (1998).

\bibitem{Peskin}  C.S. Peskin, G.M. Odell and G.F. Oster, Biophysical
Journal, 65, 316 (1993).
\end{references}
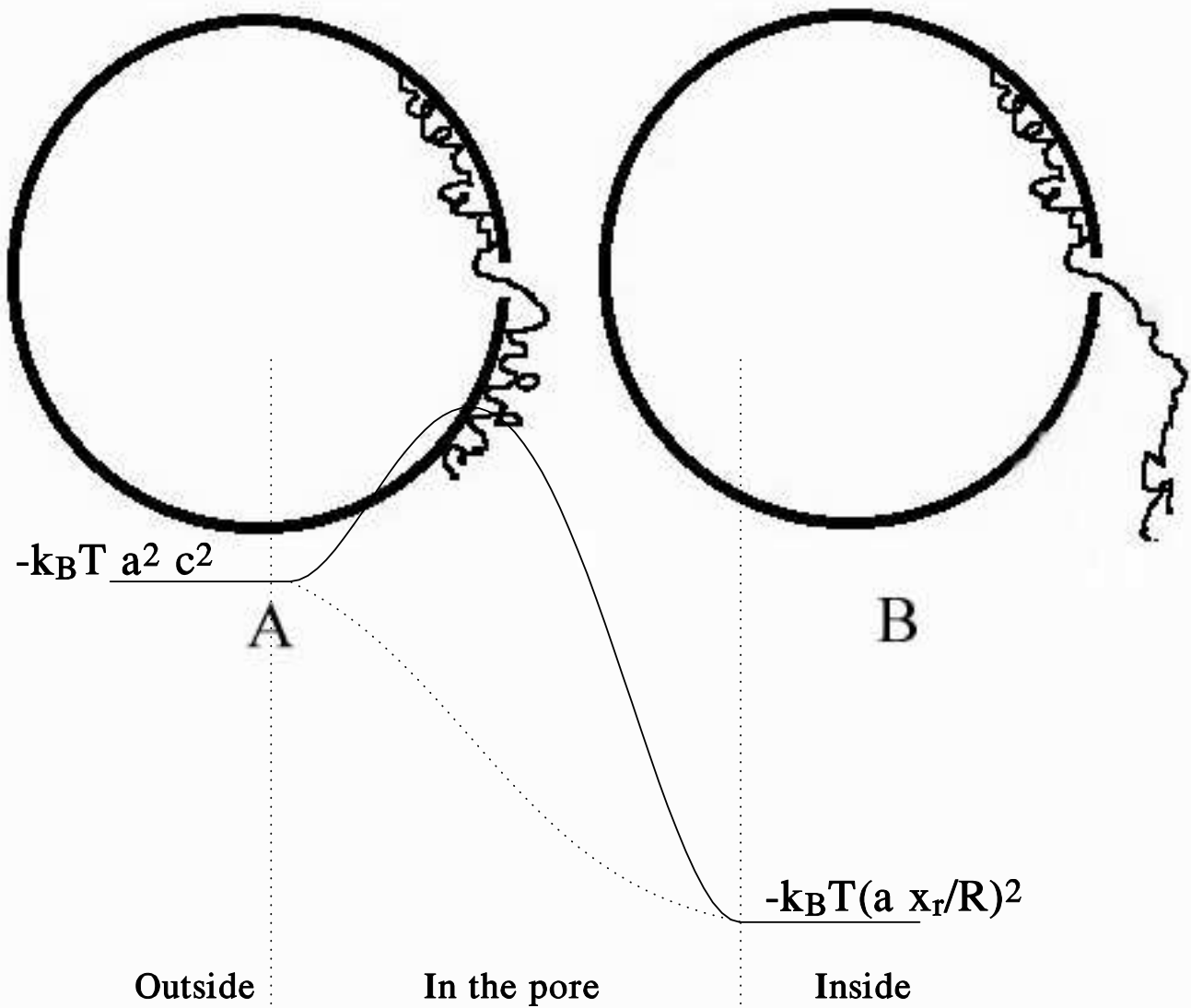
\end{document}